\documentclass[12pt]{article}
\usepackage{pazha}
\usepackage{graphicx}
\usepackage{lscape}
\usepackage{amssymb}
\usepackage{amsmath}
\usepackage{graphicx}
\graphicspath{{E:/!Gravitational_Lensing/!Images/Diploma/}}
\usepackage[T2A]{fontenc}
\tightenlines
\def\ÐÍ{$\pm$}

\begin{document}

{\it Will be published in ``Astronomy Letters'', 2011, v.37, N7, pp. 483-490}

\bigskip

\title{\bf Observations of Lensed Relativistic Jets as a Tool of Constraining Lens
Galaxy Parameters}

\author{\bf T.I.Larchenkova\affilmark{1,*}, N.S.Lyskova\affilmark{2} and A.A.Lutovinov\affilmark{3,**}}

\affil{ {\it Astro Space Center, Lebedev Physical Institute, Russian
Academy of Sciences, Profsoyuznaya str., 84/32, Moscow, 117997
Russia}$^1$\\
{\it Moscow Institute of Physics and Technology, Institutskii per. 9, Dolgoprudnyi, Moscow obl., 141700 Russia}$^2$\\
{\it Space Research Institute, Russian Academy of
Sciences, Profsoyuznaya str., 84/32, Moscow, 117997 Russia}$^3$}
\vspace{2mm}
\received{25 January 2011}

\sloppypar
\vspace{2mm}
\noindent

The possibility of using lensed relativistic jets on very small angular
scales to construct proper models of spiral lens galaxies and to
independently determine the Hubble constant is considered. The system
B0218+357 is used as an example to illustrate that there exists a great
choice of model parameters adequately reproducing its observed large-scale
properties but leading to a significant spread in the Hubble constant. The
jet image position angle is suggested as an additional parameter that allows
the range of models under consideration to be limited. It is shown that the
models for which the jet image position angles differ by at least $40^o$ can
be distinguished between themselves during observations on very small
angular scales. The possibility of observing the geometric properties of
lensed relativistic jets and measuring the superluminal velocities of knot
images on time scales of several months with very long baseline space
interferometers is discussed.

\vspace{10mm}

\vfill

$^*$ e-mail: tanya@lukash.asc.rssi.ru

$^{**}$ e-mail: aal@iki.rssi.ru

\clearpage

\section*{INTRODUCTION}

\vskip -5pt

Investigating the physical properties and formation mechanisms of
large-scale relativistic jets from active galactic nuclei, radio galaxies,
and quasars is one of the most important problems in modern astrophysics.
In particular, studying the structure and dynamics of the central regions of
extragalactic synchrotron sources, including the study of regions with sizes
of the order of several Schwarzschild radii in which relativistic particles
observed as extended jets are ejected (Meier 2009), will be a priority
direction of research with very long baseline space interferometers
(Kardashev 2009) planned to be launched in the immediate future.

The gravitational lensing of galactic nuclei, quasars, and compact regions
of radio galaxies with largescale relativistic jets allows these distant
astrophysical objects to be observed even now (Nair et al. 1993; Patnaik et
al. 1993, 1995). In the immediate future, as the resolution of the
instruments used increases, it will allow individual features of their jets,
for example, the counterjet unobservable in the absence of lensing, to be
also studied. Observing the lensed images of a relativistic jet whose knots
move with superluminal velocities due to a small angle between the line of
sight and the jet direction gives an unique opportunity to measure the
velocity of such bright knots in shorter time intervals than for unlensed
jets.

Apart from the questions related to the investigation of physical properties
of the jets themselves in gravitationally lensed systems, the questions
related to the possibility of constraining the model parameters describing
the surface density distribution in the lens and its position relative to
the emission source are also interesting. Limiting the number of models, in
turn, will allow the most important cosmological parameter, the Hubble
constant, to be more properly estimated from observations of the most
compact gravitationally lensed systems in which the position of the spiral
lens galaxy cannot be determined with present-day instruments. The source
B0218+357 ($z = 0.96$; Patnaik et al. 1993, 1995), for which the time delay
between the compact core images has been measured with a good accuracy
(Biggs et al. 1999; Cohen et al. 2000), but there is an uncertainty in
determining the relative positions of the lens and the source (York et
al. 2005), is an example of such a gravitationally lensed system.

It should be noted that the lensed jet images retain the geometric shape of
the jet on scales of tens of microarcseconds ($\mu as$), i.e., the spatial
structure in the images on such scales does not change, but, depending on
the model of a gravitationally lensed system, the jet image position angles
change (see, e.g., Larchenkova et al. 2011), whose values can be used as an
additional parameter in modeling. Observations with space interferometers on
very small angular scales (tens of $\mu as$) will allow this parameter to be
measured. This will lead to a limitation of the number of possible models
and, as a result, to a more accurate determination of the Hubble constant.

In this paper, which is a logical continuation of our previous paper
(Larchenkova et al. 2011), we use the source B0218+357 as an example to
consider the problems announced above and the approaches to their
solution. The questions related to the choice of models for the mass
distribution and position of the lens galaxy that allow the observed
large-scale picture of lensing to be adequately reproduced, in particular,
the separation between the source's compact core images, the intensity ratio
of these images, and the ringlike structures, are discussed in Section 1.
We show that it is necessary to take into account the finite width of the
relativistic jet when comparing its modeled images with observational
data. The conditions under which the position angles of the jet images on
very small angular scales can be measured with an accuracy high enough to
constrain the model parameters for the gravitationally lensed system being
studied are investigated in Section 2. In Section 3, we present our
calculations of the velocity of bright knots in the jet images depending on
the model of the lens mass surface density distribution and compare the
derived visibility functions. Our results are briefly discussed in the last
section.

\section{ALLOWANCE FOR THE FINITE JET
WIDTH AND COMPARISON WITH VLA
OBSERVATIONS}

Previously (Larchenkova et al. 2011), we modeled the multiple images of a
jet emerging when it is lensed by a spiral galaxy whose surface density was
specified by three multicomponent models reflecting the structure of spiral
galaxies. We showed that within the framework of the proposed models for the
gravitationally lensed system Â0218+357 there exists a fairly wide choice of
model parameters adequately reproducing its observed properties (the
intensity ratio of the compact source's images A and B $I_A/I_B \simeq
(3.1-3.7)$, the image separation $\simeq 335$ $mas$, and the direction of
the large-scale jet) but leading to a significant spread in the Hubble
constant. In contrast, the semiring structure observed in the radio band and
produced by the lensing of the large-scale jet appears only for a limited
set of jet parameters and direction. The lens galaxy is represented as a
singular disk truncated at a characteristic distance $a_d$ from the center
and placed in an isothermal halo with a characteristic size $a_h$ (Keeton
and Kochanek 1998).  In this case, the contribution from dark matter to the
rotation curve at distances $R < a_d$ is taken into account by introducing a
coefficient $f_d$ equal to 0.85 (Sackett 1997). The lensing potential, the
lens equation, and the expressions for the image magnification and basic
parameters (the ellipsoid inclination $i$, the ellipsoid axis ratio for the
disk $q_{3d}$ and halo $q_{3h}$) used in this model are given in Larchenkova
et al. (2011) (Model I).

To illustrate the aforesaid, we chose three sets of model parameters
adequately reproducing the observed properties of the source B0218+357,
including the presence of a semiring structure:

$$i = 0^o, q_{3h} = 0.8, q_{3d} = 0.5, a_d = 1.0,$$
$$i = 0^o, q_{3h} = 0.6, q_{3d} = 0.45, a_d = 1.0,$$
$$i = 10^o, q_{3h} = 0.6, q_{3d} = 0.05, a_d = 1.0,$$ 

which we will designate below as I, II, and III, respectively. The quantity
$a_d$ is expressed in units of the Einstein-Chwolson radius.

Previously (Larchenkova et al. 2011), we used the approximation of an
infinitely thin jet when modeling gravitationally lensed relativistic jets,
i.e., the relativistic jet was represented as an infinitely thin segment
with a constant intensity at each of its points.  However, to properly
compare the observational data with the modeling results, it is important to
take into account the finite jet width at least for two reasons.  First,
when an extended source crosses the caustic curve, the magnification of the
emerging images is not infinite but depends both on the size of the source
itself and the intensity distribution over it and on the properties of the
lensing potential (Schneider et al.  1999). Second, the observations are
performed with a finite angular resolution, as a result of which any fine
details of the images turn out to be blurred.

\begin{figure*}
\begin{center}
\includegraphics[width=0.8\textwidth]{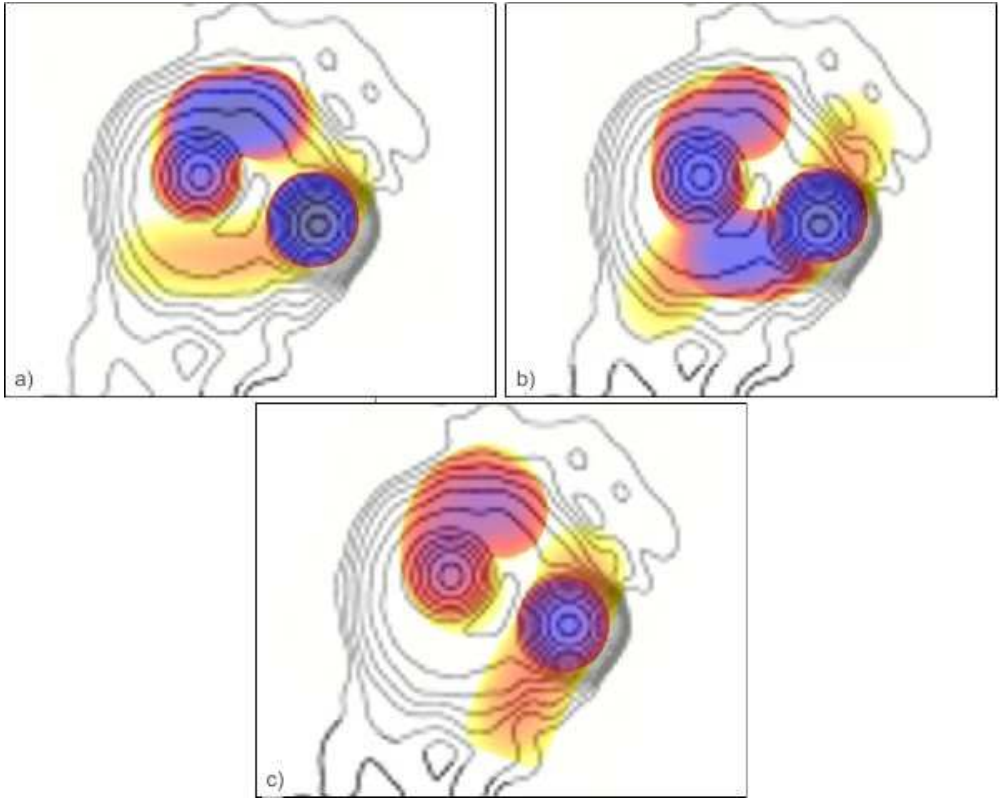}
\caption{\rm Comparison of the results of modeling the relativistic jet in B0218+357 (blue corresponds to the maximum intensity, yellow to the minimum) with the results of
VLA observations at 15 GHz taken from Biggs et al. (1999) (indicated by contours) for the sets of parameters I (a), II (b), and
III (c).}
\end{center}
\end{figure*}

The parameters of the jet itself, such as the diameter, the position angle,
and the intensity distribution along it, the presence and spatial positions
of bright knots in the lensing region, are not known for sure for the system
under consideration. Therefore, below we use the observed (and rescaled in
accordance with the distance) parameters of the nearest and best studied
relativistic jet in the source M87 (Sparks et al. 1996; Matveyenko and
Seleznev 2011) to compare the modeled and observed large-scale pictures of
lensing. In particular, at a distance corresponding to the neighborhood of
the caustic crossing by the jet for the three sets of model parameters for
the system B0218+357 under consideration, the jet diameter is assumed to be
$\simeq 0.02$ Einstein-Chwolson radius ($\simeq 46$ pc or $3.3\times10^{-3}$
arcsec). At the same time, it should be noted that near the nozzle, i.e.,
the ejection region of the relativistic plasma flow, the jet width is
probably $\simeq 0.05$ pc (see, e.g., Matveyenko et al.  2010). For the
source B0218+357, this corresponds to $\simeq 3.6\times 10^{-6}$ arcsec.

To compare the modeled large-scale picture of gravitational lensing for the
source Â0218+357 with VLA observations (Biggs et al. 1999), the resolution
of this instrument should be taken into account. It is 0.12 arcsec at 15
GHz, which exceeds the jet width in the caustic crossing region by several
tens of times.  Thus, the bright extended arcs emerging when the jet crosses
the caustic curves during observations with the above resolution turn out to
be significantly blurred in both intensity and spatial scale.

Figure 1 shows the results of our modeling for the above three sets of
parameters performed by taking into account the VLA angular resolution and
the maximum magnification near the caustic crossing by an extended (in both
length and width) jet with a constant intensity at each of its points
(Schneider and Weiss 1986; Keeton et al. 2003) against the background of the
VLA map for Â0218+357 at 15 GHz (Biggs et al. 1999). Since the distributions
of the intensity and bright knots along the jet (which are observed in
large-scale relativistic jets and whose flux is lower than the observed core
flux only by several times; see, e.g., Sparks et al. (1996) for M87) for the
system under consideration are unknown, we performed our modeling by
assuming the absence of bright knots whose images will also be blurred and
will not change significantly the large-scale jet image.

We see from Fig. 1 that, apart from quantitative relations between the
compact core images, extended structures that to a certain extent correspond
to the observed ones can be obtained for all sets of parameters under
consideration. In this case, different values of the Hubble constant are
obtained for different sets: $H_0(I) = 35.8$ km s$^{-1}$ Mpc$^{-1}$,
$H_0(II) = 46.3$ km s$^{-1}$ Mpc$^{-1}$, $H_0(III) = 68.8$ km s$^{-1}$
Mpc$^{-1}$.  Thus, additional observed parameters should be introduced to
properly determine its value. The position angle of the jet image near the
nozzle accessible to observation with very long baseline space
interferometers can be one of such parameters.

\section{POSITION ANGLES OF THE JET IMAGES
NEAR THE EJECTOR OF RELATIVISTIC
PARTICLES}

According to the VLBA survey of extragalactic sources (Kovalev et al. 2005)
for Â0218+357 on maximum projected baselines of this ground-based
interferometer ($440 \times 10^6$ wavelengths), whose angular resolution is
comparable to the VLBI one, the correlated flux density does not decrease to
zero but is about 200 mJy. Therefore, observations of this source with very
long baseline space interferometers seem interesting from the viewpoint of
determining the structure and direction of the relativistic jet at very
small angular distances from the nozzle (the ejector of relativistic
particles). Let us consider the possibility of using such observations of
the jet images to refine the model of a gravitationally lensed system
followed by the refinement of the Hubble constant.


\begin{figure*}
\begin{center}
\includegraphics[width=1.0\textwidth]{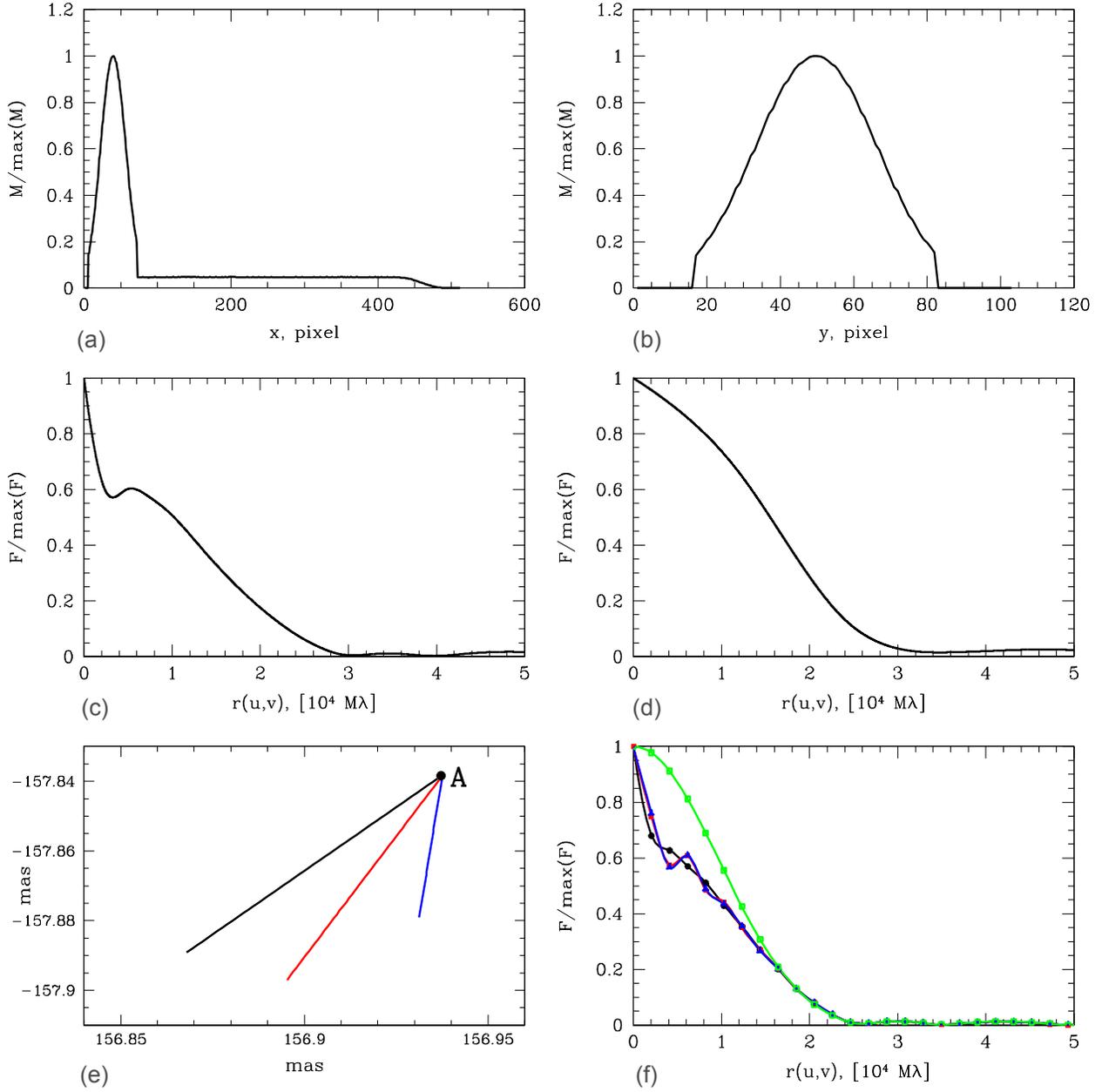}
\label{jetiki}
\caption{\rm Intensity distributions along (the X axis) and across (the Y axis) the jet images (a), (b) and the corresponding visibility
functions (c), (d). The initial phase (30 $\mu as$) of the relativistic jet emerging from image A (e) and the corresponding visibility
functions (f) for the three sets of model parameters: I (black line), II (red line), and III (blue line). The coordinates from the
lens position in mas are indicated along the axes. For comparison, the green line indicates the image visibility function
for a jet directed at an angle of $90^o$ eastward to model III.}
\vfill
\end{center}
\end{figure*}

As the limiting cases, Figs. 2a and 2b show the intensity distributions
along and across one of the jet images (the set of parameters I). The length
of the unlensed jet is 30 $\mu as$ from the nozzle (while the length of the
image itself turns out to be larger due to the gravitational lensing
effect); the intensity distribution across the jet was specified by a
Gaussian function with FWHM equal to 0.1 pc ($\simeq 7\times 10^6$ arcsec);
one pixel is equal to one $\mu as$. Figures 2c and 2d display the
corresponding visibility functions, where the normalized image intensity and
the radius vector in the UV-plane in wavelengths are plotted along the
vertical and horizontal axes, respectively.

Figure 2e shows the position angle of the jet emerging from the brighter
compact core image (image A) for the three sets of model parameters
considered.  The black, red, and blue lines indicate, respectively, the
model with the set of parameters I, for which the jet image position angle
is $135^o$ (note that the position angle of the line connecting images A and
B is $67^o$; Biggs et al. 1999), the model with the set of parameters II and
a position angle of $155^o$, and the model with the set of parameters III
and a position angle of $178^o$. For the convenience of comparison, the
initial points of all jets (the position of image A for the compact source)
were brought into coincidence at the point corresponding to the position of
image A for the set of parameters I. Figure 2f displays the corresponding
visibility functions for three jet and nozzle images, which are actually
intermediate cases between those presented in Figs. 2c and 2d. We see that
for a given angular resolution (10 $\mu as$) a change in the jet direction
by $\simeq 40^o$ causes changes in the corresponding visibility
functions. The possibility of recording such changes will depend on the
signal-to-noise ratio. For comparison, the green line in Fig. 2f indicates
the visibility function for a jet with a position angle of $-
90^o$. Summarizing the aforesaid, we conclude that the models for which the
jet image position angles differ by more than $40^o$ can be distinguished
between themselves during observations with very long baseline
interferometers.

\section{THE VELOCITIES OF KNOTS IN THE JET
IMAGES}

One more possibility to refine the model parameters for a gravitationally
lensed system on very small angular scales is to measure the velocity of the
image for a bright jet knot. The superluminal motions of relativistic jets
were also measured from the change in the positions of bright knots with
time (see, e.g., Jorstad et al. 2001). In the case of gravitational lensing
of a jet with a bright knot giving rise to its multiple images, the brighter
knot image will move with a velocity higher than both the velocity of the
unlensed knot and the velocity of the fainter image.  Thus, when such a knot
is present in the lensed jet, its displacement can be measured in a shorter
time interval than for the unlensed jet.

As the studies showed, the velocity of the knot image depends on the chosen
lens model. For example, for a jet segment with a length of 200 $\mu as$
from the nozzle in the case of B0218+357, the sets of parameters I, II, and
III give, respectively, velocities along the image of the jet emerging from
image A that are a factor of 2.9, 2.4, and 1.4 higher than the velocity
$v_0$ directly along the jet (to be more precise, along the jet projected
onto the plane of the sky). Note that the velocities along the fainter image
B turn out to be considerably (by several times) lower in all three cases.

Given the velocity with which the knot image moves along the jet and the
resolution of the instrument used in the observations, it is easy to
determine the time interval after which it is necessary to observe the
source to measure this motion. For example, if the velocity along the jet is
of the order of the speed of light, then the time interval needed to record
the knot displacement at the interferometer's angular resolution of $\simeq
10$ $\mu as$ for the model with the set of parameters I is no more than one
month.  


\begin{figure*}
\begin{center}
\includegraphics[width=0.65\textwidth]{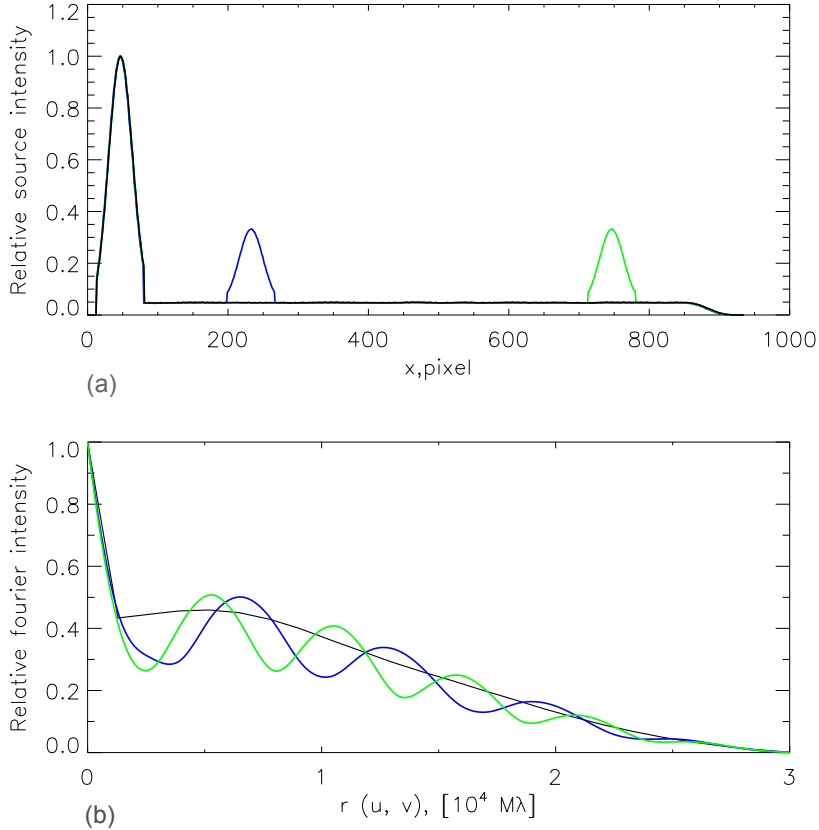}
\label{knot}
\caption{\rm Intensity distribution (a) and visibility functions (b) for the relativistic jet emerging from image A for the set of
parameters I in the following cases: at a uniform intensity (black line), in the presence of a knot with an intensity of 0.3$I_A$
at a distance of $\simeq$ 16 $\mu as$ (blue line), and in the presence of a knot with an intensity of 0.3$I_A$ at a distance of $\simeq$ 50 $\mu as$ (green line).}
\end{center}
\end{figure*}


Since it is not known in advance whether there are bright knots at the
angular distance from the nozzle under consideration, the question of
whether the visibility functions will be distinguished in the presence of
such a bright knot at a certain distance from the nozzle and in its absence
becomes interesting. For the set of parameters I and the jet parameters
given in the preceding section, we chose three possible realizations for our
analysis: (1) a jet with a core and without a bright knot, (2) a jet with a
core and a bright knot located at an angular distance of 16 $\mu as$ and
with an intensity that is a factor of 3 lower than the core one, and (3) a
jet with a core and a bright knot located at an angular distance of 50 $\mu
as$ and with the same intensity as that in case 2. Figure 3 shows the
intensity distribution along the jet image and the visibility functions for
the three chosen realizations.

We see that the visibility functions for the cases considered in the range
of projected space interferometer baselines from $5 \times 10^9$ to $3
\times 10^{10}$ wavelengths differ significantly from one another and these
differences can be recorded at an appropriate signal-to-noise ratio.

\section{CONCLUSIONS}

Because of small angular separations between the lensed source images,
systems where the lens is a spiral galaxy, for example, Â0218+357, require
modeling to determine the relative positions of the source and the lens,
without knowing which, in turn, the Hubble constant cannot be
determined. Thus, determining the Hubble constant turns out to be model
dependent.  This necessitates the introduction of additional model
parameters that can be derived from observations on very small angular
scales with the goal of limiting the number of possible models. The position
angle of the jet images emerging from a region close to the central energy
engine and, in the presence of bright knots in this spatial jet region,
their velocity in the brightest jet images can be such parameters for
gravitationally lensed systems with relativistic jets.

Having modeled the images of the large-scale relativistic jet in the source
B0218+357 emerging when it is lensed by a spiral galaxy whose surface
density distribution is described by a disk and a softened halo placed in a
singular isothermal dark matter halo, we chose several sets of model
parameters for the mass distribution and relative positions of the lens and
the source that adequately reproduce the observed large-scale picture of
lensing. Not only the separation between the source's compact core images
and the intensity ratio of these images but also the extended ringlike
structures observed on angular scales of $0.3$ arcsec (Biggs et al. 1999)
are reproduced for the chosen set of parameters. When compared with the VLA
observational data at 15 GHz (Biggs et al.  1999), the modeling results for
a fixed relativistic jet diameter with allowance made for the instrument¡¯s
angular resolution show good agreement for all of the selected model
parameters. Different values of the Hubble constant and different jet image
position angles and knot image velocities along the jet correspond to
different sets of parameters.

Based on the constructed visibility function for the chosen set of model
parameters, we showed using the system B0218+357 as an example that the
models for which the jet image position angles differ by at least $40^o$ can
be distinguished between themselves during observations on very small
angular scales (tens of $\mu as$).

Assuming the existence of a bright knot in a region close to the ejector of
relativistic particles, we calculated the velocity of the knot image
depending both on the velocity of the unlensed knot and on the model of the
lens mass surface density distribution, along with the geometry of the
gravitationally lensed system, and constructed the visibility functions for
a jet without a knot and with a knot located at different distances from the
nozzle. The visibility functions for the cases considered were shown to
differ from one another in the range of projected space interferometer
baselines from $5 \times 10^9$ to $3 \times 10^{10}$ wavelengths.

In addition, for jets with superluminal motions using the sets of parameters
considered here, the velocity of the knot image can be measured during
observations with an angular resolution of $\sim 10$ $\mu as$ by comparing
the observational data obtained with a time interval of one month. Thus, by
measuring the displacement of individual bright knots in the images with
time, it becomes possible to measure the propagation velocity of the jet
within the framework of a specified model for a gravitationally lensed
system.

\section{ACKNOWLEDGMENTS}

We wish to thank L.I. Matveyenko, V.A. Demichev, and N.S. Kardashev for
discussions and useful remarks.  This work was supported by the ``Origin,
Structure, and Evolution of Objects in the Universe'' Program of the
Presidium of the Russian Academy of Sciences, the Program for Support of
Leading Scientific Schools (project no. NSh-5069.2010.2), and the State
contracts P1336 and 14.740.11.0611.


\pagebreak


\end{document}